\documentclass[aps,prl,twocolumn,nopacs,superscriptaddress]{revtex4}
\usepackage[utf8]{inputenc}

\usepackage{graphicx}  
\usepackage{dcolumn}   
\usepackage{bm}        
\usepackage{amssymb}   
\usepackage{amsmath}
\usepackage{units}
\usepackage[english]{babel}
\usepackage[dvipsnames]{xcolor}
\usepackage{xfrac}

\usepackage[%
colorlinks=true,
urlcolor=RoyalBlue,
linkcolor=RoyalBlue,
citecolor=RoyalBlue,
]{hyperref}

\usepackage{sidecap}
\sidecaptionvpos{figure}{t}  

\usepackage[type1]{libertine}                                        
\usepackage{textcomp}
\usepackage[scaled=.85]{beramono}
\usepackage[libertine,cmintegrals,cmbraces,vvarbb,slantedGreek]{newtxmath}
\usepackage[scr=boondoxo]{mathalfa}
\usepackage{bm}
\usepackage[lf]{carlito}

\usepackage{braket}
\usepackage{bm}

\makeatletter
\DeclareRobustCommand{\cev}[1]{%
  \mathpalette\do@cev{#1}%
}
\newcommand{\do@cev}[2]{%
  \fix@cev{#1}{+}%
  \reflectbox{$\m@th#1\vec{\reflectbox{$\fix@cev{#1}{-}\m@th#1#2\fix@cev{#1}{+}$}}$}%
  \fix@cev{#1}{-}%
}
\newcommand{\fix@cev}[2]{%
  \ifx#1\displaystyle
    \mkern#23mu
  \else
    \ifx#1\textstyle
      \mkern#23mu
    \else
      \ifx#1\scriptstyle
        \mkern#22mu
      \else
        \mkern#22mu
      \fi
    \fi
  \fi
}
\makeatother

\begin{document}

\title{Surface reconstruction induced anisotropic energy landscape of bismuth monomers and dimers
on the Si(001) surface}

\author{Haonan Huang}
\email[Corresponding author; present affiliation: Princeton university; electronic address:\ ]{hnhuang@princeton.edu}
\affiliation{Max-Planck-Institut f\"ur Festk\"orperforschung, Heisenbergstraße 1,
70569 Stuttgart, Germany}
\author{J. Christian Schön}
\affiliation{Max-Planck-Institut f\"ur Festk\"orperforschung, Heisenbergstraße 1,
70569 Stuttgart, Germany}
\author{Christian R. Ast}
\affiliation{Max-Planck-Institut f\"ur Festk\"orperforschung, Heisenbergstraße 1,
70569 Stuttgart, Germany}
\date{\today}

\begin{abstract}
Spin qubits have attracted tremendous attention in the effort of building quantum computers over the years. Natural atomic scale candidates are group-V dopants in silicon, not only showing ultra-long lifetimes but also being compatible with current semiconductor technology. Nevertheless, bulk dopants are difficult to move with atomic precision, impeding the realization of desired structures for quantum computing. A solution is to place the atom on the surface which opens possibilities for atom level manipulations using scanning tunneling microscopy (STM). For this purpose, bismuth appears to be a good candidate. Here, we use ab-initio methods to study theoretically the adsorption of bismuth atoms on the Si(001) surface and investigate the adsorption sites and the transitions between them. We demonstrate the complex influence of the dimer row surface reconstruction on the energy landscape seen by a bismuth monomer and a dimer on the surface, and find anisotropic transition paths for movement on the surface. From a deposition simulation we obtain the expected occupation of adsorption sites. Our work lays the foundation for further application of bismuth atoms as qubits on silicon surfaces.
\end{abstract}

\maketitle

\section{Introduction}
Over the past two decades, ground breaking advances have been achieved to construct quantum computers, with the promise of solving problems intractable on classical supercomputers like factorization and search algorithms \cite{feynman1982simulating,ekert1996quantum,grover1996fast,shor1999polynomial,vandersypen2001experimental,nielsen2002quantum,lanyon2007experimental,monz2016realization}. Among different architectures of quantum computers, silicon based quantum computers with group V dopant atoms, such as phosphorus, as spin-qubits are of great interest \cite{zwanenburg2013silicon,chatterjee2021semiconductor}, and various proposals \cite{Kane1998,Stoneham2003,morton2008solid,hill2015surface,lei2022simple} as well as realizations \cite{tyryshkin2012electron,muhonen2014storing,watson2017atomically,hile2018addressable,broome2018two,he2019two,fricke2021coherent} have been made. 

Bismuth, the heaviest stable group V element, is a promising substitute for phosphorus in silicon based quantum computers \cite{Mohammady2010}. The performance of bismuth in spin lifetime experiments has been confirmed by experiments to be at least as good as the one of phosphorus, while the higher nuclear spin of bismuth offers the advantage of a larger Hilbert space to operate in \cite{Morley2010,wolfowicz2013atomic,ranjan2021spatially}. Also, the large hyperfine splitting of bismuth makes hybrid nuclear-electronic qubits possible \cite{Delgado2011,Morley2013}. Additionally, bismuth can be used in combination with phosphorus to support the realization of those proposals that require two species of dopants \cite{Stoneham2003}.

Nevertheless, bismuth atoms studied till now are usually buried in bulk silicon, and it is experimentally challenging to precisely position and manipulate such bulk dopants to build desired structures, which impedes realizing quantum computing proposals relying on such a level of control \cite{Kane1998}. On the other hand, single atom evaporation on surfaces and subsequent manipulation and characterization with atomic precision by means of scanning tunneling microscopy (STM) have become a mature technology \cite{serrate2010imaging,khajetoorians2019creating,ding2021tuning,kot2022electric,liebhaber2022quantum,machida2022zeeman,schneider2021atomic,schneider2023proximity,schneider2023probing,sierda2023quantum,jolie2022creating,huda2020tuneable,huang2020quantum,huang2020tunnelling,veldman2021free,huang2021phd,singha2021engineering,huang2021spin,huang2022universal,karan2023tracking,wang2023universal,siebrecht2023microwave,karan2022superconducting,villas2021tunneling,villas2020interplay}. In light of this, bismuth atoms on silicon surfaces could be a novel controllable and flexible platform holding much promise for quantum computing. 

Despite the great potential of single adsorbed bismuth atoms, previous research of bismuth on silicon surfaces has only focused on relatively high coverage bismuth layers, both experimentally and theoretically \cite{Kawazu1981,oyama1981adsorption,fan1990growth,fan1992determination,park1994strain,Tang1994,koval1995adsorption,Wasserfall1995,pyatnitskii1996effects,qian1996structure,naitoh1997scanning,gavioli1998electronic,miki1999atomically,naitoh1999bismuth,Miki1999,bowler2000structure,Naitoh2000,naitoh2001scanning,macleod2004bismuth,Bobisch2007,belosludov2007electronic,bannani2008studies,wang2008orthogonal,javorsky2010electronic,takayama2012tunable}. Therefore, a thorough theoretical investigation of the adsorption and energy landscape of individual bismuth atoms on silicon surfaces is desirable. Since evaporation of bismuth atoms in experiments usually results in a mixture of monomers and dimers \cite{Kawazu1981,oyama1981adsorption,Brewer1996,Miki1999}, it is beneficial to study the energy landscape of both monomers and dimers on surfaces to cover all scenarios.

Among different silicon surfaces, the Si(001) surface is commonly used in industry for building circuits and devices; therefore, a system based on this surface has the potential of a smooth integration into the current silicon-based technology. In addition, the Si(001) surface is semiconducting, while the Si(111) surface has metallic surface states \cite{chadi1979si,chadi1979atomic,kageshima1992theory,yoo2002electrical,smeu2012electronic}, which may decrease spin coherence time or even quench the magnetic moment. Consequently, the Si(001) surface is better suited for quantum computing applications than the Si(111) surface, and thus we study bismuth atoms on the Si(001) surface.

Here, we present an ab-initio exploration of the energy landscape of bismuth monomers and dimers adsorbed on the Si(001) surface. We study the energy minima and the energy barriers separating them, and construct a picture of the anisotropic movement of bismuth monomers and dimers due to the interaction with the underlying reconstructed silicon dimer rows on the Si(001) surface. We also simulate the deposition process and obtain an occupation distribution of the energy minima in the absence of thermalization. Our results pave the way for further theoretical and experimental research on bismuth atoms on the Si(001) surface and possible future applications as qubits in quantum computers or building blocks in quantum simulators.

\section{Methods}

Like other group-IV elements \cite{stekolnikov2002absolute}, silicon exhibits various reconstructions of different surfaces. The Si(001) surface, in particular, has been studied extensively over the years on various levels of theory, ranging from a tight-binding description, density functional theory (DFT), multi-configuration self-consistent field (MCSCF) and configuration interaction (CI) to quantum Monte Carlo (QMC) \cite{chadi1979si,chadi1979atomic,chadi1980reexamination,zhu1989electronic,penev1999effect,kageshima1992theory,Shkrebtii1995,Ramstad1995,Healy2001,martin2020electronic}. When applying such quantum chemistry techniques, it is found essential to model the Si surface either via a large atom cluster or, alternatively and usually better, by employing a slab of finite thickness with periodic boundary conditions (PBC) in order to reproduce the correct ground state configuration of such a reconstruction.\cite{Healy2001,martin2020electronic} Thus, in this study, we employ a slab, and perform DFT calculations using the Quantum ESPRESSO suite \cite{giannozzi2009quantum} with pseudo-potentials from the SSSP library \cite{lejaeghere2016reproducibility,prandini2018precision} and a PBE functional, since the number of atoms required is quite large, and many energy calculations will be needed to map out the landscape and the transition paths of the Bi atoms.

For all calculations, the bottom surface of the slab is passivated with hydrogen atoms, which is crucial to obtain the correct results. All single point energies are calculated using a wave function cutoff energy of $60\,\text{Ry}$ while more computationally intensive calculations like NEB calculations use a cutoff energy of $30\,\text{Ry}$ for efficiency. In all structural relaxations, the lowest two silicon layers and the hydrogen atoms at the bottom are fixed in order to stabilize the structure, while all other atoms are allowed to relax in all directions. The structural relaxation threshold is always $1\times10^{-6}\,\text{Ry}$ for the energy of the total system and $1\times10^{-5}\,\text{a.u.}$ for the force on every atom allowed to move.

Clean Si(001) surface calculations use a $4\times2$ supercell (eight surface atoms) with $2\times4\times2$ k-points and five atomic layers of silicon. The $c$ length of the calculation cell is $6.55\,\text{nm}$, corresponding to a "thickness" of the vacuum between the periodically repeated slabs of ca.\ $5.9\,\text{nm}$ in the z-direction.

Similarly, calculations of the bismuth atoms on the Si(001) surface mostly employ a $4\times4$ supercell (sixteen surface atoms) to ensure a sufficient separation of bismuth atoms in neighboring (periodically repeated) cells and six atomic layers of silicon in total (for some massive deposition simulations we have also employed a $4\times2$ supercell). For calculations of a single total energy, $4\times4\times2$ k-points are used, while for the more time-consuming NEB calculations, $2\times2\times1$ k-points are used to reduce calculation load. The $c$ length of the simulation cell for the case of a monomer, i.e., a single Bi atom, is the same as for the clean Si(001) surface, while the one for the dimer is $3.56\,\text{nm}$ (corresponding to a vacuum between slabs of around $2.87\,\text{nm}$ thickness), for efficiency. For stability and efficiency, the calculations are not spin-polarized if not stated otherwise. But, for comparisons, we will perform some spin-polarized calculations and discuss the difference in the case of bismuth monomers. Convergence with respect to the number of k-points, the cutoff energy, the vacuum spacing and other calculation parameters has been tested to ensure that the above settings yield the correct energy.

\section{Results}

\subsection{Clean Si(001) surface}

\begin{figure}[ht]
    \centering
    \includegraphics[width=\columnwidth]{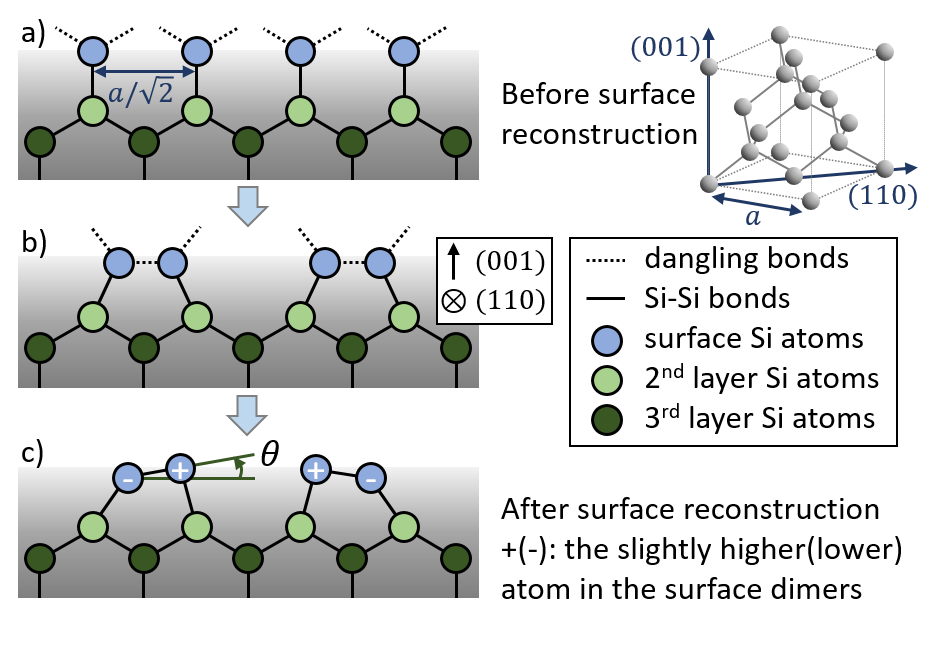}
    \caption{\textbf{Formation of Si(001) surface reconstruction.} a) Newly truncated Si(001) surface, cross section (viewed from (110) direction), showing two dangling bonds per surface atom. The distance between nearest surface atoms is $a/\sqrt{2}$, where $a$ is the lattice constant of the unit cell of bulk silicon. b) The formation of surface dimers reduces the number of dangling bonds to one per atom, reducing the energy. c) ``Buckling'' of the surface dimer, forming upper atom (denoted by +) and lower atom (labeled by -) configurations, further lowering the energy.}
    \label{fig1}
\end{figure}

\begin{figure}[ht]
    \centering
    \includegraphics[width=\columnwidth]{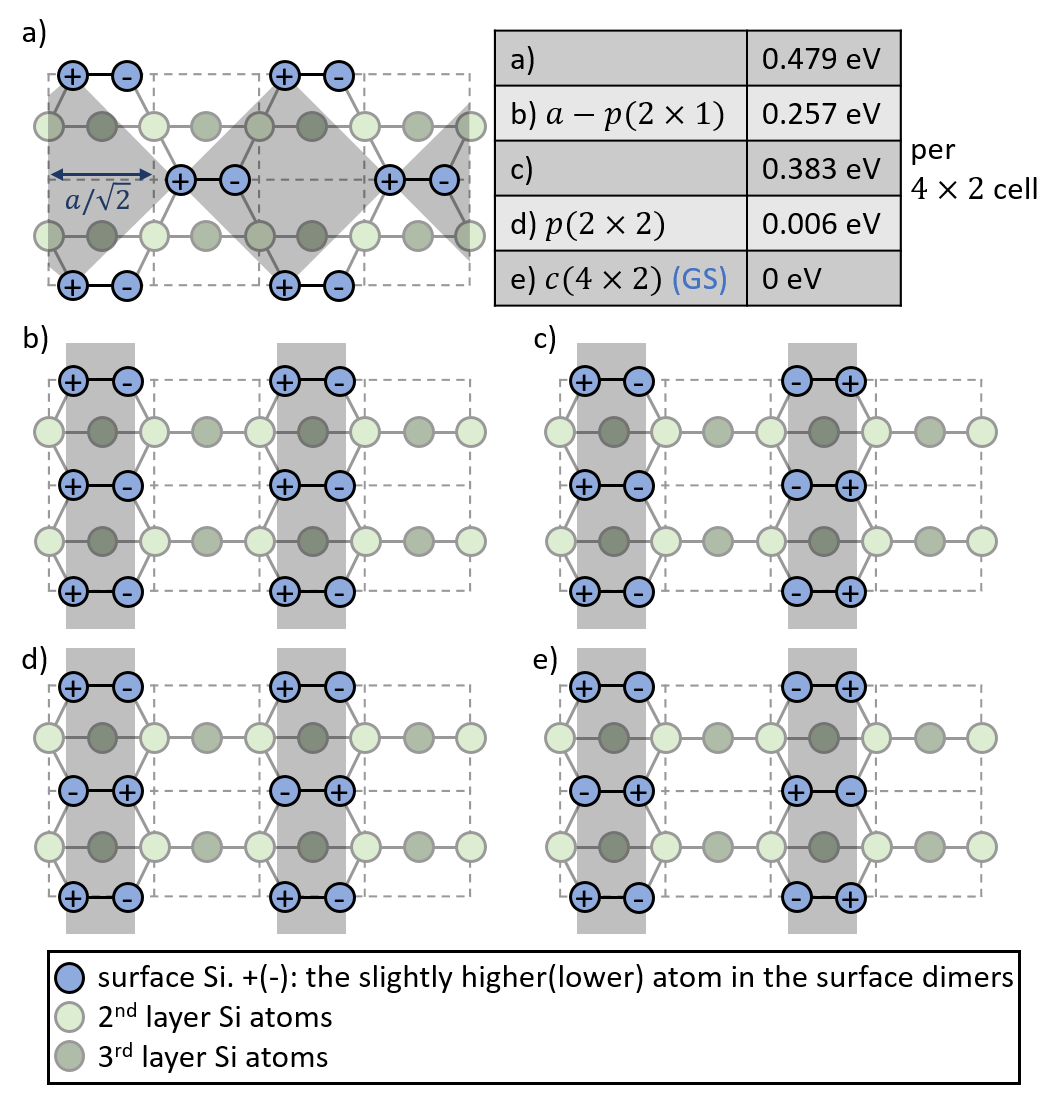}
    \caption{\textbf{Various configurations of Si(001) surface reconstructions and their corresponding energy.} All panels are shown in $4\times2$ supercell (surface atoms form a square lattice depicted with dashed grid with lattice constant $a/\sqrt{2}$, and the $4\times2$ supercell corresponds to eight surface unit cells). a) Checkerboard configuration. b-e) Dimer row reconstruction. b) All parallel, which can reduce to a $2\times1$ supercell ($a-p(2\times1)$). c) Parallel within a dimer row, opposite between adjacent rows. d) Opposite for adjacent dimers within a dimer row, parallel between dimer rows, which can reduce to a $2\times2$ supercell ($p(2\times2)$). e) Alternating buckling direction both within and in between dimer rows ($c(4\times2)$, ground state).}
    \label{fig2}
\end{figure}

On a newly truncated Si(001) surface before reconstruction, the surface atoms form a perfect square net with the lattice constant being $a/\sqrt{2}$ ($a$ is the lattice parameter of the cubic unit cell of silicon) \cite{chadi1979si} (Fig.\ \ref{fig1}). Due to truncation, each surface silicon atom has two dangling bonds, each having one electron (Fig.\ \ref{fig1}(a)). To lower the energy, neighboring surface atoms pair up forming bonds, leaving one dangling bond per surface atom (Fig.\ \ref{fig1}(b)). If this were the complete story, the newly formed surface dimers would appear parallel to the surface, and the remaining dangling bonds would form $\pi$-bonds resulting in a conducting surface state \cite{chadi1979atomic,chadi1979si,kageshima1992theory,martin2020electronic}. However, what actually happens is that the dimers experience a further tilt (``buckle'') and charge is depleted at the lower atom and transferred to the higher atom, forming a lone pair on the higher atom, which results in an insulating surface \cite{Healy2001,martin2020electronic} (Fig.\ \ref{fig1}(c)). In this study, we denote the higher (lower) atom in the surface dimer as $+(-)$. 

In general, there are two ways how all surface atoms can pair up into dimers without generating defects (or mixtures of these two ways): By forming a checkerboard pattern (Fig.\ \ref{fig2}(a)) or by producing dimer rows (Figs.\ \ref{fig2}(b-e)). For a checkerboard pattern, it turns out that the only stable configuration is the one where all dimers are in the same buckling orientation (Fig.\ \ref{fig2}(a)), and this pattern is energetically less favorable than the dimer rows (Fig.\ \ref{fig2}). 

For a dimer row reconstruction, the surface is divided into dimer rows which protrude away from the average surface in the z-direction (gray shaded area in Fig.\ \ref{fig2}) and deeper-lying valleys in-between. Depending on whether adjacent dimers within one dimer row and in the neighbor dimer rows feature identical or opposite buckling orientations, there are various possibilities (Figs.\ \ref{fig2}(b-e)). We find that it is energetically preferred that the adjacent dimers within one row buckle in opposite orientations forming a staggered arrangement. The energy minimum (ground state, GS) is the $c(4\times 2)$ configuration, where adjacent dimer rows also buckle in different directions (Fig.\ \ref{fig2}(e)), although it is lower in energy than the $p(2\times 2)$ configuration (Fig.\ \ref{fig2}(d)) only by a marginal amount. This indicates that at elevated temperatures or under a sufficiently strong tunneling current in STM, the dimer rows will flip-flop between $c(4\times 2)$ and $p(2\times 2)$ phases. As a consequence, on typical observation time scales, the buckling averages out and the  asymmetry of the surface dimers is not observed in experiments \cite{kageshima1992theory,Shkrebtii1995,Hata2001,Healy2001}. The above findings are consistent with the  literature \cite{Shkrebtii1995,Ramstad1995,Healy2001,martin2020electronic}. In the following we will ignore such dynamical aspects and use the $c(4\times 2)$ configuration as the basis for further calculations involving bismuth.

\subsection{Bismuth monomer on Si(001)}

\subsubsection{Local minima and corresponding energy} 
\begin{figure}[ht]
    \centering
    \includegraphics[width=\columnwidth]{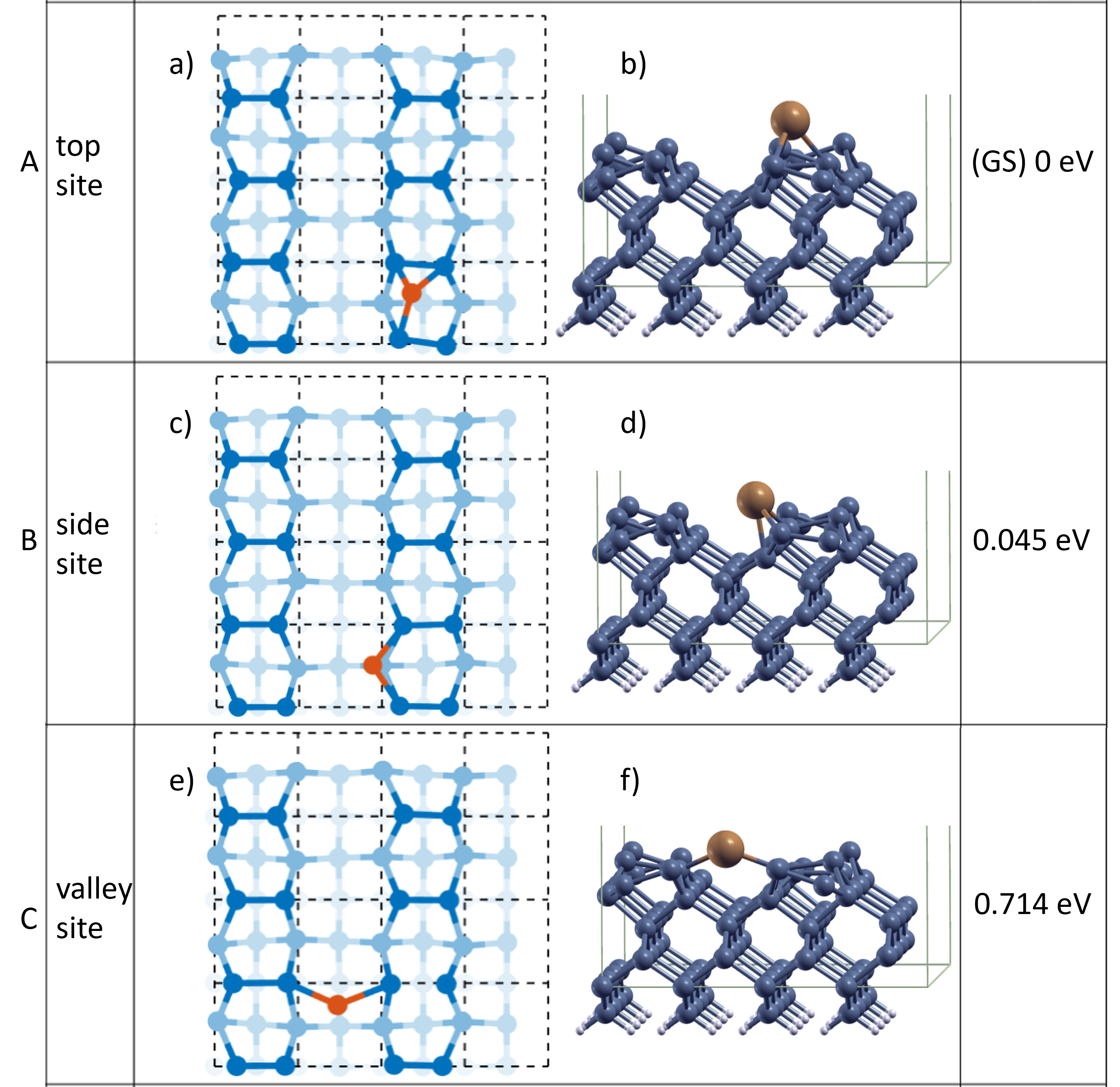}
    \caption{\textbf{Local minima of Bi monomers.} a-b) top site A. c-d) side site B. e-f) valley site C. a,c,e) are top views and b,d,f) are 3D views of the atomic structure. The corresponding energy with respect to the ground state (GS) is shown on the right hand side.}
    \label{fig3}
\end{figure}

We first relax a single bismuth atom from many random positions just above the Si(001) surface and find three stable candidates for adsorption sites, one on top of the dimer rows (``top site'', A, Figs.\ \ref{fig3}(a,b)), one on the side of the dimer rows (``side site'', B, Figs.\ \ref{fig3}(c,d)) and one in the valley between the dimer rows (``valley site'', C, Figs.\ \ref{fig3}(e,f)).

\begin{figure}[ht]
    \centering
    \includegraphics[width=0.9\columnwidth]{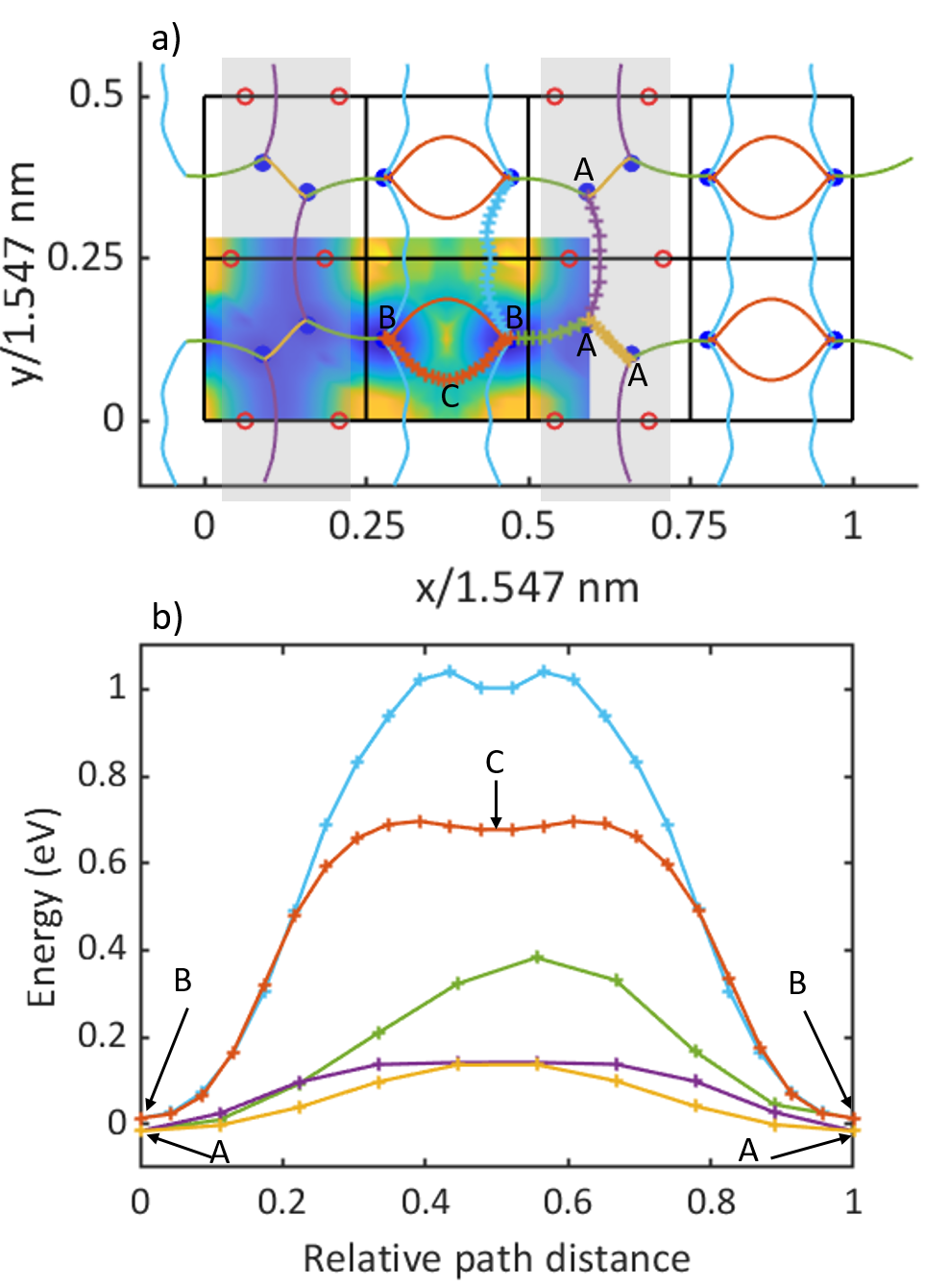}
    \caption{\textbf{Energy landscape, transition paths and energy barriers of Bi monomers.} a) The blue disks are the important local minima positions of a bismuth monomer (A: top site and B: side site. C: valley site is not plotted with blue disks). The red circles denote the positions of surface silicon atoms. The optimal transition paths between the local minima are drawn as lines of different color in between. The global energy landscape is depicted as a color plot in the lower left part of the $4\times2$ supercell. b) The energy barrier along the transition paths shown in panel a), with the same coloring. The energy barrier along the top of the dimer rows is much smaller than the one perpendicular to the dimer rows. Notice that the dip in the middle of the light blue curve might correspond to another local minimum, but we ignore it in our discussion due to its high energy, low barrier and tiny basin of attraction.}
    \label{fig4}
\end{figure}

The Bi atom on the top site (A) sits between adjacent Si surface dimers near the middle, but closer to the lower (-) atom of the Si dimer. According to the symmetry of the $c(4\times 2)$ surface, there are two equivalent sites between adjacent Si dimers and, therefore, eight in total in the $4\times 2$ supercell (Fig.\ \ref{fig4}). This configuration is the ground state with an adsorption energy of $4.647\,\text{eV}$.

The Bi atom on the side site (B) sits on the edge of the dimer row bonding with only two surface Si atoms from adjacent Si dimers. There are eight such sites in the $4\times 2$ supercell due to symmetry (Fig.\ \ref{fig4}). Those adsorption configurations have nearly mirror symmetry, but the mirror symmetry is not exact due to the buckling of the silicon dimers. The side site has around $0.045\,\text{eV}$ more total energy than the top site.

The Bi atom on the valley site (C) sits in the middle of the valley and has significantly higher energy, rendering it a kinetically stable but thermodynamically unstable adsorption site. Later we will show that the energy barrier from the valley site to the side site is quite low, proving that its kinetic stability is also quite low. Molecular dynamics simulation at room temperature confirms that the Bi atom at the valley site quickly moves into the side site. Therefore the only relevant stable sites at room temperature are the top and the side site.

\subsubsection{Transition paths and energy barriers between minima} After obtaining the adsorption sites, it is of interest to study the transition paths and corresponding energy barriers between these local minima. It is not only crucial to study the transition path between nearest monomer sites, but also along or perpendicular to the Si dimer rows, because this provides an estimate to which extent bismuth atoms are mobile on the surface in various directions. Here, we use the nudged elastic band (NEB) method, where the (locally) optimal minimum energy path is found automatically starting from a given initial path between two local minima; here, one must keep in mind the limitations of the NEB approach, i.e., we usually only find the locally optimal path that is closest to the initially proposed path. 

The resulting paths are shown in Fig.\ \ref{fig4}(a) and the total energy along the path is shown in Fig.\ \ref{fig4}(b). In Fig.\ \ref{fig4}(a), the red circles denote the surface silicon atoms and the shaded rectangles show the dimer rows along the y-axis. The blue disks are the monomer adsorption sites including eight top sites (labeled with ``A'') and eight side sites (labeled with ``B''), C denotes the valley sites, as discussed before.

We have identified five irreducible paths to navigate across the whole energy landscape: between nearest top sites (A-A yellow), between two top sites crossing one silicon dimer (A-A magenta), between top site and closest side site (A-B green), between side sites across the valley going through a valley site (B-C-B orange) and between side sites along the valley (B-B light blue). Combining this with the translation symmetry of the surface, these paths span the energy landscape and form a network (Fig.\ \ref{fig4}(a)).

The energy barrier between adjacent top sites (A-A yellow) is the lowest, around $0.153\,\text{eV}$, and that between top sites along the dimer row (A-A magenta) is only slightly higher, around $0.158\,\text{eV}$, although the path is much longer and thus less probable in reality due to an entropic barrier; for a general discussion of entropic barriers, we refer to the literature\cite{Schoen03a,Neelamraju2017a}. The energy barrier between adjacent top and side site (A-B green) is around $0.399\,\text{eV}$ staring from top site. The barrier between side sites across the valley (B-C-B orange) is significantly higher, around $0.683\,\text{eV}$, and that along the valley (B-B light blue) is the highest, around $1.026\,\text{eV}$.

One important piece of information here is the energy barrier from the valley site C to the side site B, which is around $0.02\,\text{eV}$. This is lower than the thermal energy at room temperature, so we expect the valley site to be unstable at room temperature.

The above finding indicates that the migration of a bismuth atom on the surface is highly anisotropic, much more difficult perpendicular to the dimer rows compared to along the dimer rows, and easier on the dimer rows than inside the valley. Also, all of the energy barriers are non-negligible even at room temperature where the thermal energy $k_\text{B}T$ is around $0.026\,\text{eV}$, and thus bismuth monomers are expected to be rather immobile on the Si(001) surface even up to room temperature. This is quite unlike to the situation for many other single atoms on surfaces which usually need rather low temperatures to survive aggregation into dimers or clusters \cite{kot2022electric}; clearly, this is an important property  of Bi atoms on the Si(001) surface, which is beneficial for device fabrication.

\subsubsection{Global energy landscape} Although the system is three-dimensional, the effective region of interest is the surface which reduces essentially to a two-dimensional problem. In this case, it is possible to directly visualize the global energy landscape of a bismuth monomer on Si(001). To obtain such a landscape representation, we divide the surface into a fine grid and relax a bismuth atom from each point on the grid. The bismuth atom starts from a certain height above the surface and is allowed to relax only in $z$ until the force in $z$ on the atom approaches zero, while all other atoms relax as usual (see methods section). Then we plot the resulting total energy as a function of the projection in the $x-y$ plane as a color plot in Fig.\ \ref{fig4}(a).

Using this method, not only the locations of the local minima but also the transition paths can be better understood. There is a low energy channel on top of the dimer rows, which protrudes slightly into the valley between the rows yielding the side site B. The energy in the valley is generally high and flat, forming a significant barrier for bismuth atoms moving perpendicular to the dimer row orientation. There is a shallow minimum in the valley near the valley site C, but with a quite sizeable basin of attraction. The energy landscape is highly anisotropic due to the directional dimer row reconstruction.

\subsubsection{Occurrence of minima from deposition} In the high temperature limit, the distribution of the occupancy of the sites will be thermal according to the energies of the minima. At room temperature, the Boltzmann factor for the top and valley sites 
\begin{equation}
\frac{p_\text{valley}}{p_\text{top}}=e^{-\frac{E_\text{valley}-E_\text{top}}{k_\text{B}T}}
\end{equation}
is around $1.0\times10^{-12}$ and that for the top and side sites 
\begin{equation}
\frac{p_\text{side}}{p_\text{top}}=e^{-\frac{E_\text{side}-E_\text{top}}{k_\text{B}T}}
\end{equation}
is around $0.18$, meaning that the thermal equilibrium neglecting the energy barrier will result in the majority being top sites, a small portion being side sites and nearly no valley sites at all.  

In the opposite limit where the temperature is insufficient for crossing the barrier, a random deposition of bismuth single atoms will result in a certain distribution of the occupancy of the sites, which can be either simulated from relaxation starting at random initial positions, or from the area of the basin in the obtained global energy landscape in Fig.\ \ref{fig4}(a). Considering the frequency, with which the minima are reached when starting from the systematic grid point approximation of the global energy landscape, the top sites, side sites and valley sites are expected to occur with about 43\%, 39\% and 18\% probability, respectively. When performing direct relaxation simulations from 402 random initial positions, we find that 39.0\% of the relaxed configurations were top sites, 46.5\% were side sites and 14.5\% were valley sites, respectively. These results are consistent within a few percent given the finite sampling size of the depositions as well as the finite resolution of the global energy landscape. At intermediate temperatures, the bismuth monomers on a valley site will move into the nearest side site; as a consequence, the majority of the occupied sites will actually be side sites, contrary to the thermal equilibrium case. Finally, if the deposition happens at very low temperatures instead, the valley sites will co-exist with the other two sites according to the percentages presented above.

\subsection{Bismuth dimer on Si(001)}
\subsubsection{Local minima: energy and structures} 
\begin{figure}[ht]
    \centering
    \includegraphics[width=\columnwidth]{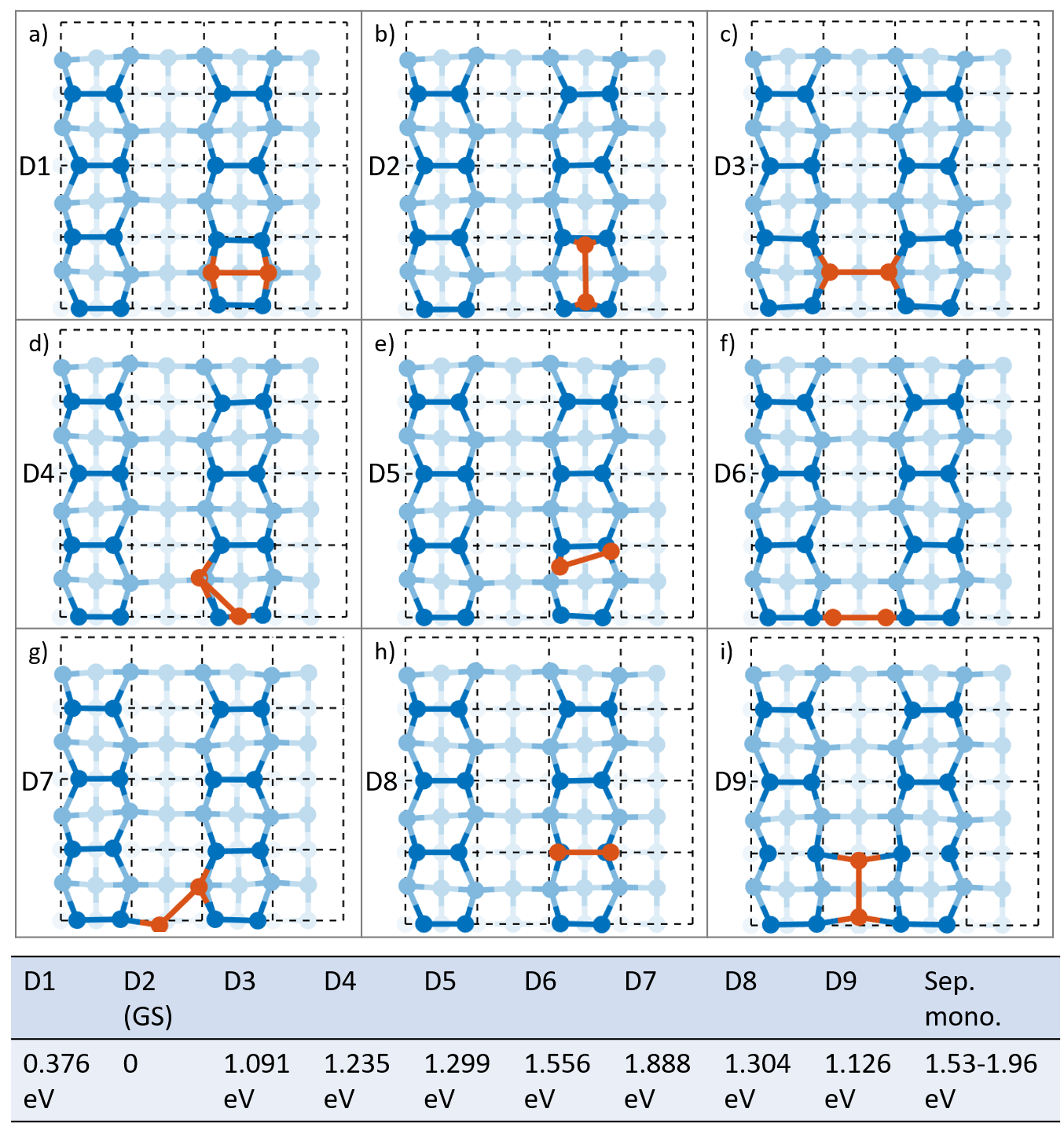}
    \caption{\textbf{Local minima of Bi dimers.} (a-i) nine dimer configurations D1-D9. The energy of the minimum is shown at the bottom, with the ground state (GS, D2) as reference. }
    \label{fig5}
\end{figure}

Going from monomer to dimer, the number of degrees of freedom is doubled, and the energy landscape becomes much more complex. One trivial possibility is that the two atoms are located in two essentially independent monomer sites that are far away from each other, such that no bonding will take place between the atoms. However, they can still interact through the underlying surface because an adsorbed atom modifies the silicon surface reconstruction in its vicinity, resulting in a range of total energies depending on the distance between the two adsorption sites. Disregarding these arrangements, we focus on those initial locations where a direct (chemical) interaction between the two Bi atoms is present. For these true dimers on the surface, we have identified nine local minima labeled D1-D9 as shown in Fig. \ref{fig5}, and computed their energy.

D1-D9 can be categorized into two location categories: on top of the Si dimer rows (D1, D2, D4, D5, D8) and in the valley between the Si dimer rows (D3, D6, D7, D9). Each category contains one configuration with the bismuth dimer along $y$ (D2 and D9), two along $x$ (D1 and D3 between surface Si dimers, D8 and D6 in line with a surface Si dimer), and tilted ones (D4, D5 and D7).

The ground state configuration is D2 with an adsoprtion energy of around $3.92\,$eV with respect to a bismuth dimer in the gas phase, where the bismuth dimer sits on top of the silicon dimer row along the y-direction (Fig.\ \ref{fig5}(b)), parallel to the Si dimer row. Rotating the bismuth dimer in D2 by 90° results in D1 (oriented in the x-direction) and a total energy around $0.376\,\text{eV}$ higher, which is the second energetically favorable configuration (Fig.\ \ref{fig5}(a)). In both cases, due to the bonding with the bismuth dimer, the buckling of neighboring silicon dimers vanishes. 

Other local minima (D3-D9) are more than $1\,\text{eV}$ higher in energy than the ground state D2, and are therefore expected to be observed much less frequently in the experiment than D1 and D2 if we only consider the thermal equilibrium distribution at room temperature. Nevertheless, as already shown in the case of monomers, the energy barriers for the Bi atoms to move on the surface are non-negligible, thus influencing the equilibration processes and the expected or observed occupation distribution at low temperatures on finite time scales. Therefore it is insightful to study the transition paths and energy barriers between these local minima for the case of the dimer.

\subsubsection{Transition paths and energy barriers between minima} 
\begin{figure*}[ht]
    \centering
    \includegraphics[width=0.95\textwidth]{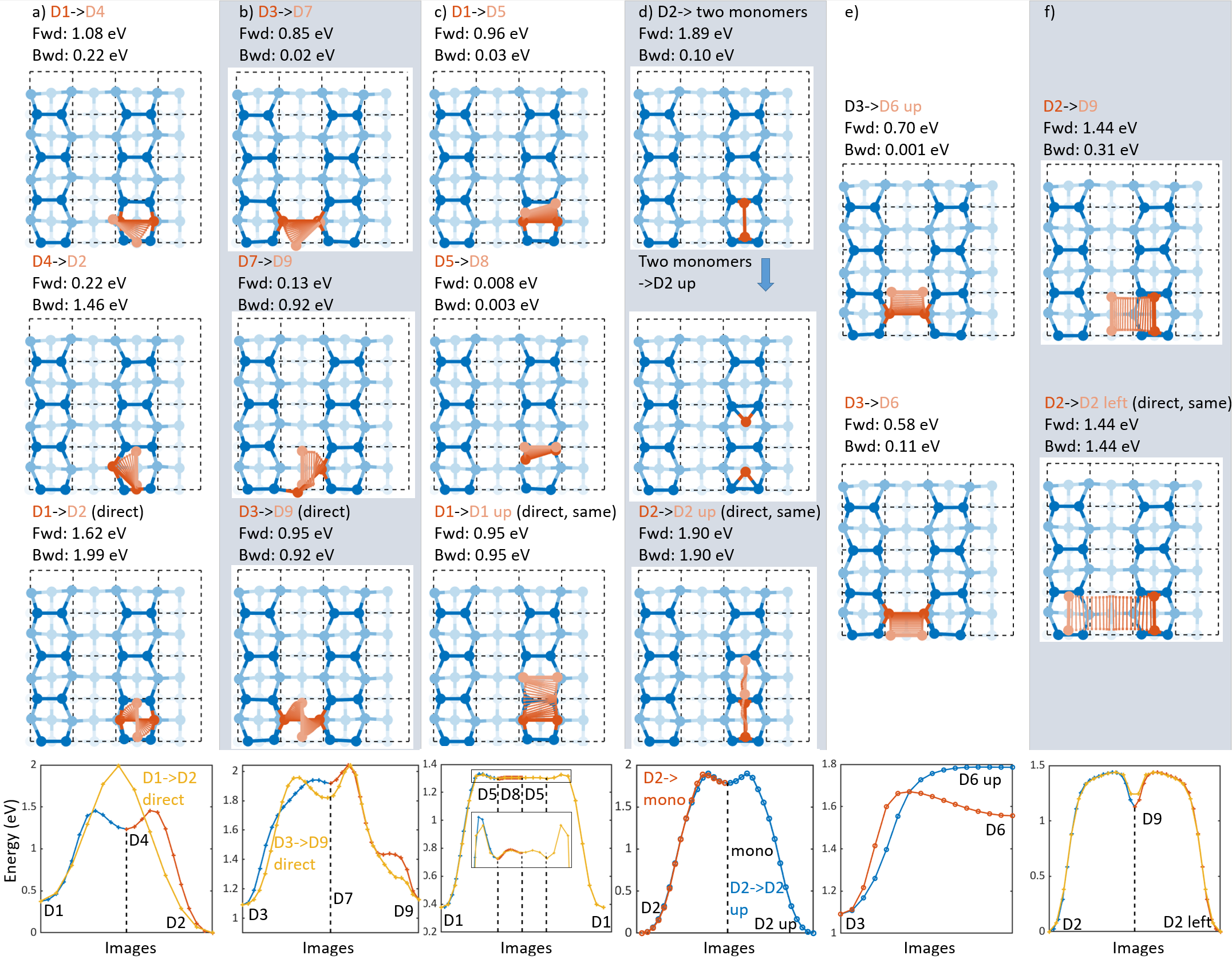}
    \caption{\textbf{Transition paths and energy barriers of Bi dimers.} (a) D1-(D4)-D2 (b) D3-(D7)-D9 (c) D1-D5-D8-D5-D1 (d) D2-separated monomers-D2 (e) D3-D6 (or D6 up) (f) D2-D9-D2.}
    \label{fig6}
\end{figure*}

To connect the minima D1-D9, there are generally three types of basic transitions and their combinations which are important: rotation (Fig.\ \ref{fig6}(a-b)), translation along $y$ (along the dimer rows, Fig.\ \ref{fig6}(c-e)) and translation along $x$ (perpendicular to the dimer rows, Fig.\ \ref{fig6}(f)).

A rotation connects configurations in approximately the same position with different orientations in the xy-plane. For Bi dimers on the dimer row, this corresponds to the path D1-(D4)-D2. If we conduct a NEB calculation directly from D1 to D2, the algorithm will identify a path with a forward barrier of 1.62\,eV. However, when including D4 as an intermediate image (such that the dimer rotates from D1 to D4 first, and then to D2), the forward barrier is lowered to 1.08\,eV. This indicates that the optimum transition path between D1 and D2 is through D4 (Fig.\ \ref{fig6}(a)). As for Bi dimers in the valley, due to the plateau feature of the energy landscape in the valley, there are many possibilities for rotations, and the algorithm can converge to several paths with similar barriers, one of them shown in the yellow curve in Fig.\ \ref{fig6}(b). Forcing the transition through D7 changes the path but does not optimize the height of the energy barrier. The energy barrier from D7 to D3 is quite low (around $0.024\,$eV), indicating the unstable character of D7. Such an instability is also reflected in the high energy of D7, which is even higher than the energy of most instances of two separated Bi atoms. 

Regarding translations, one scenario is to move along the dimer rows (along the $y$-axis). For a dimer at D1 to move up to the next D1 position, NEB calculations automatically gives the optimal path through D5 and D8 (D1-D5-D8-D5-D1), shown in Fig.\ \ref{fig6}(c). In the energy barrier plot at the bottom, it can be seen that D5 and D8 are local minima, but both with high energies and very small barriers stabilizing them, making these shallow minima very unstable at room temperature and very likely to move into the D1 configuration. For a dimer at D2 to move up to the next D2 configuration, the calculation shows that the bismuth dimer will disintegrate into two monomers as an intermediate step, then recombine forming a dimer in the new position (Fig.\ \ref{fig6}(d)). For a dimer at D3 to move down or up, it will pass through a D6-type configuration on either route (Fig.\ \ref{fig6}(e)). One interesting observation here is that there exist two different D6-type minima, labeled D6 and D6-up, which have different energies and different barriers. The reason is that D6 connects two upper (+) atoms in a silicon surface dimer while D6-up, which is one dimer shift up from D6, connects two lower (-) atoms in a silicon surface dimer, resulting in such an asymmetry. We note that the D6 minimum is much more stable than the D6-up configuration.

Another type of translation is the movement perpendicular to the dimer rows, connecting Bi dimers with the same orientation in different locations. For example, for a dimer at D2 to move perpendicular to the dimer rows, it will pass though D9 in the middle of the valley (Fig.\ \ref{fig6}(f)).

As a consequence, at room temperature, D5, D6, D7, D8 might be unstable and thus transform into other local minima. A dimer at D5 or D8 will move into D1, while one at D7 or D6 will move into D3. Coincidentally, these four configurations have the highest energies among all dimer configurations (c.f. Fig.\ \ref{fig5}). To realize, and stabilize, these configurations, very low temperatures are necessary.

\subsubsection{Deposition simulation}
\textbf{General comments.} At room temperature, the Boltzmann factor of the energy difference between the ground state D2 and the second lowest energy state (D1) is $4.8\times10^{-7}$ indicating that the thermal equilibrium distribution of the occupancy of the configurations will consist nearly solely of D2. Nevertheless, due to non-negligible energy barriers, bismuth dimers can remain partially in the initial relaxed configuration after deposition. It is highly non-trivial to compute the exact distribution of occurrence of the various configurations, at finite temperature, on finite time scales, by modeling the time evolution of the system via the transitions through barriers for a multitude of possible initial states; this is beyond the scope of this study. Here, we only consider the opposite limit where no thermalization after deposition takes place (low temperature limit), i.e., we only allow a relaxation into the nearest configuration, but no subsequent transitions.

To simulate the deposition process, two alternative routes exist. One is to deposit two atoms separately one after the other (after the full relaxation of the first atom), and another is to deposit a pre-formed bismuth dimer, which we will discuss in the next two subsections. Intermediate cases such as the deposition of the second Bi atom while the first one is still in the process of relaxation, or allowing for the possibility of the Bi-atom to bounce off the surface and landing at another site due to the accumulation of kinetic energy and momentum while approaching the attractive potential of the surface from the gas phase, are not considered here.

\begin{figure}[ht]
    \centering
    \includegraphics[width=\columnwidth]{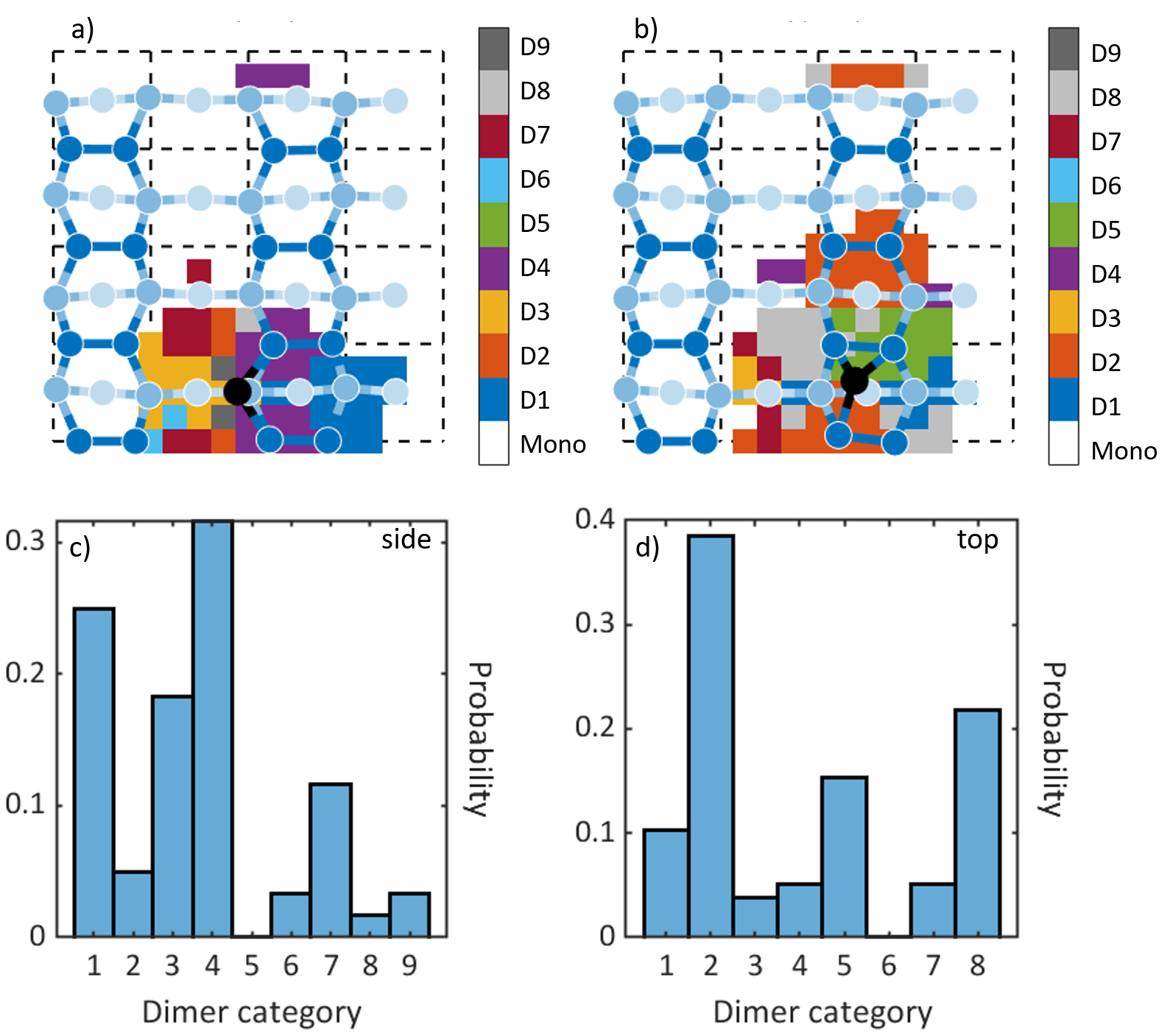}
    \caption{\textbf{Deposition of a second bismuth atom on a surface with a relaxed bismuth monomer already present.} a,b) Dependence of the relaxed dimer configurations on the initial position of the second atom, starting from a side (a) or a top (b) site monomer. c,d) Probability distribution of the relaxed dimer configuration starting from a side (c) or a top (d) site monomer. It can be seen that some dimer configurations do not form by adding two monomers separately.}
    \label{fig7}
\end{figure}

\textbf{Deposition simulation I: two single bismuth atoms one at a time.} 
In the case of two separate depositions, we assume no correlation on the initial positions. Thus, unless the second atom is close enough to the relaxed position of the first one to spontaneously form a dimer, they will remain separate and no dimer is produced. Nevertheless, this simulation corresponds to the situation of limited dimer formation with only monomers present in the gas phase during deposition and can indicate the range of distances from the initially deposited first Bi atom, which allows the formation of a bismuth dimer; the maximal distance can be denoted the formation radius of the dimer.

We only consider the first monomer in the side site (Fig.\ \ref{fig7}(a)) or in the top site (Fig.\ \ref{fig7}(b)) because the valley site is unstable, shown as a black disk for clarity. Starting from these two initial positions, the second bismuth atom is relaxed from a grid of initial positions covering the surface at a certain distance (in the z-direction) from the surface; this grid is analogous to the one we employed to generate the representation of the global landscape for a single Bi atom on the Si surface. Both bismuth atoms are allowed to freely relax and all other atoms also relax as discussed in the methods section. The resulting configurations after full relaxation are categorized and plotted in Fig.\ \ref{fig7}.

The first thing to notice is that dimers only form when the second atom is deposited within a small area around the first monomer as expected. This area has a radius on the order of the lattice constant of Si, which means that the formation distance of the dimer is short without thermal activation. In this area, the distribution of dimer configurations is highly anisotropic. 

Starting with a Bi atom on the side site (Fig.\ \ref{fig7}(a)), the distribution has approximate mirror symmetry (not exact due to buckling of the Si dimers) with the mirror plane through the side site along $x$. The final result depends largely on the relative orientation of the initial position of the second atom with respect to the first site: if it is close to the $x$ direction, D3 or D1 will be produced depending on whether the second atom starts in the valley or on the dimer row. If it is tilted, D4 or D7 will be generated. A histogram of probabilities is shown in Fig.\ \ref{fig7}(c), which shows that the majority of the dimers belongs to D1, D3, D4 and D7.

If the first Bi atom is on the top site (Fig.\ \ref{fig7}(b)), the symmetry is less pronounced than for the side site case. Nevertheless, the dimer formation's outcome with respect to the orientation is similar. In particular, if the line connecting the initial positions of the two monomers runs roughly parallel to the $y-$direction, the dimer D2, which is parallel to $y$, is produced. A histogram of probabilities is shown in Fig.\ \ref{fig7}(d), which shows that the majority of the dimers found belong to D2, D5 and D8; thus, the distributions for the side site and top site cases are complementary, with D6 and D9 only rarely appearing.

\textbf{Deposition simulation II: one dimer.} A more natural pathway to obtain bismuth dimers is to deposit pre-formed bismuth dimers from the gas phase, which we simulate in the following. The complexity here comes from the orientation of the dimer, which can point in arbitrary directions in space. To account for this, we consider a spherical coordinate system with polar angle $\theta$ (angle with respect to the $z$-axis) and azimuthal angle $\phi$ (angle between the projection of the dimer into the $xy$ plane and the $x$-axis). Taking the inversion symmetry of bismuth dimers into account, we can roughly sample the whole spherical space of initial orientations by selecting the following thirteen directions:
\begin{itemize}
    \item \#1 (along $z$-direction): $\theta=0^{\circ}$
    \item \#2-9 ($45^{\circ}$ to $z$-direction): $\theta=45^{\circ}$, \\
    $\phi=90^{\circ},45^{\circ},0^{\circ},-45^{\circ},-90^{\circ},-135^{\circ},-180^{\circ},-225^{\circ}$
    \item \#10-13 (in $x-y$ plane): $\theta=90^{\circ}$, $\phi=90^{\circ},45^{\circ},0^{\circ},-45^{\circ}$.
\end{itemize}

\begin{figure*}[ht]
    \centering
    \includegraphics[width=0.85\textwidth]{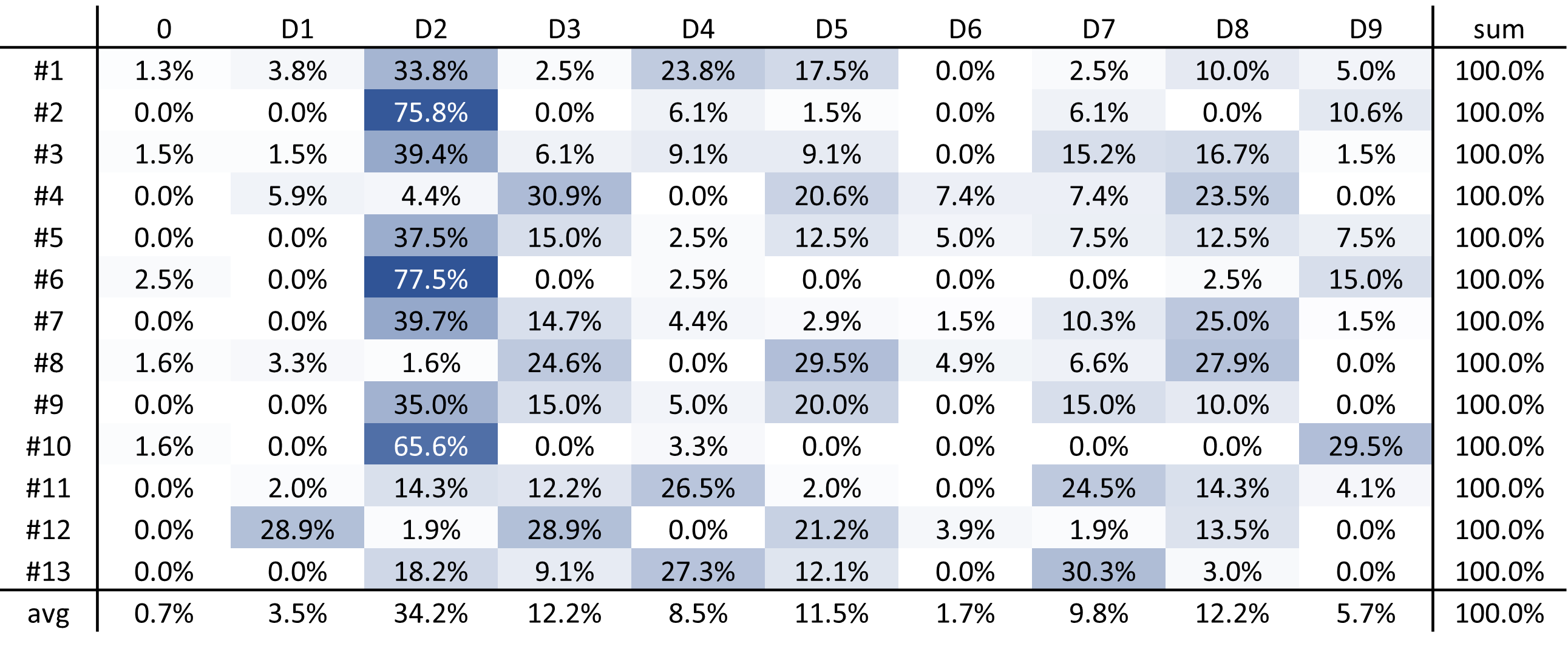}
    \caption{\textbf{Deposition of a bismuth dimer.} \#1-\#13 are 13 initial orientations of the bismuth dimer in the gas phase, and D1-D9 are the relaxed configurations of the bismuth dimer on the surface. 0 indicates the break-up of the dimer into two separated monomers. No transitions were allowed, i.e., the temperature was assumed to be zero.}
    \label{fig8}
\end{figure*}

Again, we can classify these initial orientations by their projection in the $xy$ plane: no projection (\#1), along $y$ (\#2,6,10), along $x$ (\#4,8,12) and tilted (\#3,5,7,9,11,13).

For each initial dimer direction, we relax from a number of random initial positions at a certain height above the surface and classify the final relaxed configuration in Fig.\ \ref{fig8}. 

One observation is that the resulting configuration distribution depends critically on the starting orientation. For D2 and D9, where the bismuth dimer is along the $y-$direction, they occur most often when the projection of the initial orientation points along the $y$ (\#2,6,10), less so for tilted initial orientation (\#3,5,7,9,11,13) and suppressed for initial orientation along $x$ (\#4,8,12). For D1, D3, D6, D8 where the bismuth dimer is along the $x$ direction, they occur most prominently when the projection of the initial orientation points along $x$ (\#4,8,12). Therefore, the orientation of a bismuth dimer in the relaxed configuration is largely correlated with the initial orientation of the dimer in the gas phase.

For a thermal deposition in the experiment, the orientation of the dimer is completely random, so only the average of \#1-13 is important. At the low-temperature limit where no thermalization occurs, D2-D5 and D7-D8 are the major minimum configurations observed. Here, we note that D1 only rarely  occurs, even though its energy is the second lowest one among all dimers; the reason for this is the small phase space of initial gas phase orientations (just case \#12) that results in D1. Furthermore, in about 1\% of the cases, the dimer split into two separate Bi atoms on the surface. At room temperature, since D5-D8 are unstable and easily transform into D1 and D3 as discussed above, the majority of the species left will belong to D1-D4. We note that the distributions for both cases considered, i.e., $T = 0$ and low temperatures below the energy barriers that separate the various minima, are far from thermal equilibrium, for which we would expect D2 being essentially the only minimum present.

\section{Discussion}
\textbf{Spin features.} The calculations performed are non-spin-polarized calculations to reduce the computational effort. Nevertheless, not only are spins especially interesting for quantum computing, but can also lead to a correction to the total energy. 

For a bismuth monomer, since bismuth has a $6s^26p^3$ outer electron configuration and thus an odd number of electrons, there must be at least one unpaired electron. Since the Si(001) surface is non-conducting, the unpaired electron will give rise to a non-zero localized spin. Indeed, calculations taking the spin into account yields spin-$1/2$ for bismuth monomers in all three adsorption sites on Si(001). To correctly reproduce the spin-$1/2$ feature, we need a supercell at least as large as $8\times8$ (four times as large as the one shown in Fig.\ \ref{fig3}) to sufficiently separate the PBC images of bismuth, requiring much more computational power than the non-spin case. This indicates a long spin-spin interaction radius of the bismuth monomers on the surface, which is beneficial for connecting qubits. 

As for the energy shift, the inclusion of the spin does not change the ground state being the top site, but the energy difference between the side and top site changes to around $67\,\text{meV}$ and that between the valley and the top site changes to around $0.74\,\text{eV}$, both just a bit higher ($20-30\,\text{meV}$) than the calculation without spin. These small corrections to the energy barrier (Fig.\ \ref{fig4}(b)) are negligible, therefore validating our previous non-spin-polarized approximation.

A bismuth dimer, on the other hand, has an even number of electrons, and is likely to be spin zero. Indeed, we have not found any dimer configuration featuring a non-zero spin in our calculations when taking the spin explicitly into account. 

A more comprehensive investigation of the spin characteristics of all bismuth monomer and dimers including the surface-mediated spin-spin interaction and including spin orbit coupling (SOC) and on-site Coulomb repulsion is beyond the scope of this study and calls for further research. Here, we remark that another potentially interesting aspect is the site dependent g-factor and its electrical control \cite{kot2022electric}.

\textbf{Dynamics at finite temperature.} Although we have analyzed the low temperature limit and the opposite thermal equilibrium limit of bismuth atoms and dimers deposited on the Si(001) surface, the intermediate regime of finite temperature can also be of great relevance for the experiments. A detailed discussion of this case is beyond the scope of this study, but we have conducted several molecular dynamics calculations of the valley adsorption configuration for the bismuth monomer and of the D5 configuration for the bismuth dimer at $300\,\text{K}$, to gain some first insights into this issue. We find that both configurations are unstable at room temperature and quickly move into deeper energy minima nearby (the side site and the D1 configuration, respectively) within a few picoseconds. This result agrees with our expectations from the discussion of the transition paths and the energy barriers we have determined in this study.

\section{Conclusion}
In this paper, we have studied the global energy landscape of single bismuth atoms and Bi dimers on a Si(001) surface. We have identified the three and nine local minimum configurations for a bismuth monomer (top, side and valley site) and a Bi dimer (D1-D9), respectively, adsorbed on the Si(001) surface, and computed their energies. We have investigated the transition paths and energy barriers between them, and we could show that the energy landscape is anisotropic as a consequence of the Si-dimer row surface reconstruction. 

We have discussed the expected frequency of occupation of the energy minima in two opposite limits of very low and high temperatures. For high temperatures, the system is in thermal equilibrium where the minima are observed according to the Boltzmann distribution, while in the low temperature limit, the system remains in the initial state that was reached after only allowing a downhill relaxation of the deposited atoms and dimers without thermalization, i.e, the likelihood of a minimum configuration being observed is proportional to the size of the basin of attraction and does not reflect its energy. For the low temperature limit, we have conducted deposition simulations to compute this probability for each configuration, where the resulting minima occupancy distribution is found to be very different from the one expected in thermal equilibrium. Regarding a precise calculation of this distribution as function of time, temperature and initial conditions of the gas phase, i.e., gas pressure and temperature, and precise composition of the gas phase (monomers, dimers, and larger Bi clusters), further research is called for, in close collaboration with the experiment. Our work lays the foundation for further research on the presence and behavior of bismuth atoms and dimers on the Si(001) surface, and should lend support to possible future quantum computing or quantum simulation \cite{sierda2023quantum} proposals for scalable quantum circuits based on the manipulation of atoms on the level of individual atoms and dimers.

\bibliographystyle{apsrev4-1new}
\bibliography{HNLib_bibtex}

\end{document}